# Probing strong interactions in p-type few-layer WSe$_2$ by density-dependent Landau level crossing

Shuigang Xu[1][†], Liheng An[1][†], Jiangxiazi Lin[1], Zefei Wu[1], Tianyi Han[1], Gen Long[1], Yuheng He[1],

Zhi-qiang Bao[2], Fan Zhang[2][*], and Ning Wang[1][*]

[1]Department of Physics and Center for Quantum Materials, the Hong Kong University of Science and Technology, Clear Water Bay, Hong Kong, China

[2]Departement of Physics, the University of Texas at Dallas, Richardson, Texas 75080, USA

†These authors contributed equally to this work.

*Corresponding author: phwang@ust.hk (N.W.), zhang@utdallas.edu (F.Z.)

## Abstract

**Atomically thin transition metal dichalcogenides (TMDCs) such as MoS$_2$ and WSe$_2$ are emerging as a new platform for exploring many-body effects. Coulomb interactions are markedly enhanced in these materials because of the reduced screening[1] and the large Wigner-Seitz radii[2]. Although many-body excitonic effects in TMDCs have been extensively studied by optical means[3,4,5,6,7,8,9], not until recently did probing their strongly correlated electronic effects become possible in transport. Here, in *p*-type few-layer WSe$_2$ we observe highly density-dependent quantum Hall states of $\Gamma$ valley holes below 12 T, whose predominant sequences alternate between odd- and even-integers. By tilting the magnetic field to induce Landau level crossings, we show that the strong Coulomb interaction enhances the Zeeman-to-cyclotron energy ratio from 2.67 to 3.55 as the density is reduced from 5.7 to $4.0 \times 10^{12}$ cm$^{-2}$, giving rise to the even-odd alternation. Unprecedentedly, this indicates a 4.8 times enhancement of the *g*-factor over its band theory value at a density as high as $4.0 \times 10^{12}$ cm$^{-2}$. Our findings unambiguously demonstrate that *p*-type few-layer WSe$_2$ is a *superior* platform for exploring strongly correlated *electronic* phenomena, opening a new perspective for realizing the elusive Wigner crystallization at a *moderate* density.**

Electronic states of semiconductors are most often described by the conventional band theory when electrons are only weakly interacting. The exchange and correlation energies that arise from the electronic Coulomb interaction can, however, dominate the kinetic energy in the dilute doping limit[1]. In atomically thin transition metal dichalcogenides (TMDCs), interaction effects can become considerably strong because of the reduced dielectric screening in two dimensions[10, 11] and the large Wigner-Seitz radius in the materials[2]. Optical experiments in low-mobility TMDCs have already identified several significant interaction effects, such as giant bandgap renormalization[3], large exciton-binding energies[4, 5], tightly bound trions[7], and negative electronic compressibility[6]. Recent progresses in fabricating *high-mobility* TMDC devices have triggered intrinsic transport experiments[12, 13, 14, 15, 16, 17, 18], with surprising observations of the giant spin susceptibility in the $\Gamma$ valley of few-layer WSe$_2$[16] and the density-dependent quantum Hall (QH) effects in the $K$ valleys of monolayer WSe$_2$[18]. Both transport anomalies most likely arise from the strong Coulomb interaction.

The strength of Coulomb interaction effects of a two-dimensional electron gas (2DEG) is often determined by the ratio of the Coulomb energy to the kinetic energy, i.e., the Wigner-Seitz radius $r_s = 1/(\sqrt{\pi n} a_B^*)$. Here $n$ is the carrier density, $a_B^* = 4\pi\hbar^2\varepsilon\varepsilon_0/(m^* e^2)$ is the effective Bohr radius corrected by the dielectric constant $\varepsilon$ and the effective mass $m^*$, $\hbar$ is the Planck's constant, $\varepsilon_0$ is the free space permittivity, and $e$ is the electron charge. Physically, $r_s \sim m^*/\sqrt{n}$ characterizes the average inter-electron spacing tunable by $n$; a larger $r_s$ indicates stronger interaction effects. For a 2DEG, $r_s \sim 10$ is considered as strongly interacting, and $r_s \sim 38$ is the critical point for the transition from the conducting Fermi liquid to the insulating Wigner crystal[1, 19].

Naturally, the *p*-type few-layer TMDCs offer the best platform *to date* for exploring such strongly correlated physics for two important reasons. First, the effective mass of the $\Gamma$ valley[16, 20] is about twice larger than those of the $K$ and $Q$ valleys in TMDCs and a few to ten times larger than those of graphene, black phosphorus, and quantum-well 2DEGs. Second, the $\Gamma$ valley has negligibly weak spin-orbit couplings (SOC)[16] and no valley degeneracy, yielding the desired pure interaction effects at low density detectable at tilted magnetic fields. In our devices, $r_s$ is estimated to be 12.2 at the lowest density $2.5 \times 10^{12}$ cm$^{-2}$, given $\varepsilon \approx 4.15$ for h-BN[21, 22] and $m^* \approx 0.75 m_0$ for $\Gamma$-holes (see Supplementary Information). Such a $r_s$ is twice larger than that of monolayer WSe$_2$ at a similar density[18, 22]. To achieve a comparable $r_s$ in Si[23, 24], GaAs[25, 26], and AlAs[27, 28], much lower densities $\sim 10^{10}$ cm$^{-2}$ were required, which is notoriously challenging.

Taking the full advantages of the comparably large $m^*$ and the negligibly weak SOC of the $\Gamma$-valley hole carriers, here we probe the strong interaction effects in *p*-type few-layer TMDCs by measuring the Landau level (LL) crossing at tilted magnetic fields as a function of the carrier density. We observe that the predominant sequence of QH states below 12 T switches from odd integers to even integers and then back to odd integers, as the density is reduced from 5.7 to $2.5 \times 10^{12}$ cm$^{-2}$. Tilting the magnetic field to induce LL crossing, we find that $g^* m^*$, a measure of spin susceptibility, increases from $5.34 m_0$ to $7.10 m_0$ as the density decreases from 5.7 to $4.0 \times 10^{12}$ cm$^{-2}$. Remarkably, the *g*-factor enhancement over its bare value predicted by the band theory is as large as $g^*/g_0 = 4.8$ at the density of $4.0\times10^{12}$ cm$^{-2}$. Although at a still moderate density, such enhancement and its density sensitivity are more striking than in any other accessible 2DEG, clearly evidencing the strong interaction effects in atomically thin TMDCs. Our findings demonstrate a superior platform and open a new perspective for exploring strongly correlated physics in reduced dimensions.

Our high-quality WSe$_2$ devices are fabricated based on the dry transfer technique reported previously[29]. Technical details about the h-BN encapsulated structure and the selective etching technique for fabricating low-temperature Ohmic contacts to WSe$_2$ can be found in Ref.[14, 15, 16, 30]. Pd is used as the contact electrodes for accessing the $\varGamma$ valley of p-type few-layer WSe$_2$. The high-performance devices have a typical Hall mobility of ~ 4500 cm$^2$/V s (at $n$=5.7×10$^{12}$ cm$^{-2}$) and enable us to observe the QH states at cryogenic temperatures.

Our typical magneto-transport data at different hole densities are shown in Fig. 1. The density is determined from the slope of the Hall resistance in the linear region. Shubnikov-de Haas (SdH) oscillations in longitudinal resistance ($R_{xx}$) and quantized plateaus in Hall resistance ($R_{xy}$) are clearly visible. At $n = 2.5\times10^{12}$ cm$^{-2}$ in Fig. 1a, the quantized plateaus are observed predominantly at the odd-integer filling factors (FFs) such as ν = 7, 9, 11, ... with the appearance of corresponding $R_{xx}$ minima. Such an unconventional sequence, similar to the one reported previously[16], is attributed to a special ratio of the Zeeman energy to the cyclotron energy, which will be elaborated below. At $n = 3.4\times10^{12}$ cm$^{-2}$ in Fig. 1b, surprisingly, the QH states are observed predominantly at the even-integer FFs instead, such as ν = 10, 12, 14, ....

To better reveal the detailed density dependence, Figure 1c displays $R_{xx}$ plotted as a function of FFs at five different densities from 5.7 to 2.5 ×10$^{12}$ cm$^{-2}$. The FFs are calculated by $\nu = nh/eB$, where $B$ is the magnetic field. At $n$ =5.7×10$^{12}$ cm$^{-2}$, the $R_{xx}$ minima predominantly appear at odd-integer FFs such as ν = 23, 25, 27, 29, .... In sharp contrast, from $n$ =4.8 to 3.4×10$^{12}$ cm$^{-2}$, the most pronounced $R_{xx}$ minima all occur at even-integer FFs. Although weaker $R_{xx}$ minima can be observed at odd-integer FFs at relatively higher $B$ because of the Zeeman effect, the even-integer QH states predominate in these regions. As the density further decreases to 2.5×10$^{12}$ cm$^{-2}$, the main $R_{xx}$ minima switch back to odd-integer FFs such as ν = 9, 11, 13, 15, ... . Not only are these $R_{xx}$ minima in Fig. 1c consistent with the quantized $R_{xy}$ plateaus in Figs. 1a and 1b, but also these results showcase the strong density dependence of the $\varGamma$-valley QH states. (A similar phenomenon has recently been reported for $K$-valley holes in monolayer WSe$_2$[18], with a much lighter $m^*$.)

For such single-valley spin-degenerate $\varGamma$-holes, the sequence of QH states is completely determined by the competitions between the cyclotron energy $E_c$ and the Zeeman energy $E_z$ that is *dressed by interactions*. As illustrated in Fig. 3c, when $2j + 0.5 < E_z/E_c < 2j + 1.5$, with $j$ a non-negative integer, the odd-integer states predominate; on the contrary, when $2j - 0.5 < E_z/E_c < 2j + 0.5$, the even-integer states predominate. (The situation is the opposite for the monolayer case, due to the presence of one anomalous LL at the band edge of one of the two K valleys[31].) Therefore, $E_z/E_c$ needs to be determined at each density in order to understand the alternating sequences. Conventionally, the measurement of coincident angles at tilted magnetic fields has been an accurate method to determine $E_z/E_c$ in a 2DEG[32]. Thanks to the weak SOC and the large $m^*$ in the $\Gamma$ valley[16], this technique is perfectly executable in *p*-type few-layer WSe$_2$. (In monolayer WSe$_2$, the out-of-plane quantization and the substantial splitting of spin, however, prevent the determination of $E_z/E_c$ through this technique[18].) At a tilt angle $\theta$, the cyclotron energy $E_c = \hbar e B_\perp / m^*$ is directly proportional to the perpendicular component of the magnetic field $B_\perp$. By contrast, the Zeeman splitting $E_z = g^* \mu_B B_t$ scales linearly with the total magnetic field $B_t$, where $g^*$ is the *effective* Landé *g*-factor, and $\mu_B$ is the Bohr magneton. As $\theta$ increases, two LLs with opposite spins and orbital indices differing by $i$ cross at the so-called coincident angles that fulfil $E_z/E_c = g^* m^* / 2 m_0 \cos\theta = i$ with $i$ a positive integer, as sketched in Fig. 3c.

We measure the SdH oscillations in our samples at various different $\theta$ at 1.7 K. The value of $\theta$ is calibrated by the simultaneous Hall measurements in the linear region, since the total carrier density does not change with $\theta$. Fig. 2 shows the variation of $R_{xx}$ vs $B_\perp$ at various different $\theta$ for two representative densities, $n_1 = 5.7 \times 10^{12}$ cm$^{-2}$ and $n_2 = 4.0 \times 10^{12}$ cm$^{-2}$. The SdH oscillations of both cases have significant angle dependence. For the case of $n_1$ in Fig. 2a, as $\theta$ increases from 0° to 28.9°, the even-integer QH states at $\nu = 22, 24, 26, ...$ become weaker and weaker and eventually vanish, whereas the odd-integer QH states are almost unchanged. Precisely, at the coincident angle of $\theta = 28.9°$, pronounced $R_{xx}$ minima (maxima) appear at the odd- (even-) integer FFs. As $\theta$ further increases to the next coincident angle at $\theta = 48.2°$, a phase reversal[33] in SdH oscillations is observed, i.e., the $R_{xx}$ minima (maxima) gradually become maxima (minima). Likewise, we can obtain one more coincident angle at $\theta = 57.4°$ for $n_1$ and three coincident angles at 24.3°, 44.1° and 54.5° for $n_2$, as shown in Fig. 2b. Two prominent features can be disclosed by comparing the evolutions of SdH oscillations at the two densities. First, their coincident angles are completely different. Second, their switching of even-odd integer FFs at those coincident angles are completely opposite. For example, at their first coincidence angles, at $n_1$ ($n_2$) the $R_{xx}$ minima occur at odd- (even-) integer FFs. Both features imply the different values of $E_z/E_c$ at the two densities.

Based on the coincidence angles, we can accurately extract the spin susceptibility $\propto g^* m^*$ by fitting to the coincidence condition, $i \cos\theta = g^* m^* / 2 m_0$. The results are shown in Fig. 3a: $g^* m^* = 5.34 m_0$ at $n_1 = 5.7 \times 10^{12}$ cm$^{-2}$ whereas $g^* m^* = 7.10 m_0$ at $n_2 = 4.0 \times 10^{12}$ cm$^{-2}$. Moreover, we can alternatively extract $g^* m^*$ from the angle-dependent amplitudes of SdH oscillations. As evidenced in Fig. 2, the maximum oscillation amplitudes occur right at the coincidence angles. In fact, the amplitudes can be described by[34, 35] $R(\theta) \propto \cos(\frac{\pi g^* m^*}{2 m_0 \cos\theta})$. The fittings to this formula in Fig. 3b yield $g^* m^* = 5.42 m_0$ at $n_1$ and $g^* m^* = 7.08 m_0$ at $n_2$, which are in good harmony with those obtained from the coincident condition. Markedly, a *tiny reduction* of the density $n$ (or increase of $r_s$) can *substantially enhance* the spin susceptibility, clearly evidencing the strong Coulomb interaction in our system.

With the extracted values of $g^* m^*$, we further obtain that at $\theta = 0°$ the values of $E_z/E_c$ are 2.67 at $n_1$ and 3.55 at $n_2$, respectively. According to the aforementioned analysis, at $n_1$ odd-integer QH states should predominate since $2.5 < E_z/E_c < 3.5$, whereas at $n_2$ even-integer QH states should predominate since $3.5 < E_z/E_c < 4.5$. Evidently, the experimental data shown in Fig. 1c completely agree with such physical predictions based on our tilted field measurements. Fig. 3c schematically illustrates the evolution of LLs as a function of $\theta$ under a constant $B_\perp$, and the locations of $\theta = 0°$ for $n_1$ and $n_2$ are also marked. Accordingly, the first observed coincident angle occurs at $i = 3$ for $n_1$ and $i = 4$ for $n_2$; as illustrated in Figs. 4c-4d, the $n$-th LL of spin up is degenerate with the ($n$+3)-th LL of spin down at $n_1$ but with the ($n$+4)-th LL of spin down at $n_2$. Moreover, the QH states follow the sequence of $\nu = 1, 2, 3, 5, 7, 9, ...$ at $i = 3$ but $\nu = 1, 2, 3, 4, 6, 8, ...$ at $i = 4$. Our $R_{xy}$ data at each case indeed exhibit consistent QH plateaus, as shown in Figs. 4a-4b.

We note that $g^*m^*$ can also be determined by a measurement in which the $B$ field is in-plane and induces a pure Zeeman splitting between the two spin sub-bands. When the field is above the critical value[25, 26] $B_p = 2\pi\hbar^2 n/\mu_B g^* m^*$, the system is fully spin polarized producing a resistance kink. However, $B_p$ is too large to achieve in our present experiment, e.g., $B_p = 46.7$ T at $n_2$=4.0×10$^{12}$ cm$^{-2}$ by using $g^*m^* = 7.10 m_0$.

In a 2DEG, Coulomb interaction may enhance both $m^*$ and $g^*$ at low density[24, 25, 28]. To identify the key factor of the enhancement of spin susceptibility $\propto g^*m^*$, we measure the temperature-dependent SdH oscillations at various different densities to extract $m^*$ (see Supplementary Information). In our samples, we do not observe any significant deviation of $m^*$ from $0.75 m_0$ by decreasing the density from 6.6 to 2.8×10$^{12}$ cm$^{-2}$. Therefore, we conclude that the interaction-driven enhancement of $g^*m^*$ is from the exchange-induced $g$-factor enhancement. In the non-interacting limit, the $g$ factor at $\Gamma$ valley should be $g_0 = 2.0$ [16] by applying the negligibly weak SOC to the Roth's formula[36]. In the present interacting case, by using the measured value $m^* = 0.75 m_0$, we obtain $g^* = 7.1$ and 9.5 at $n_1$=5.7×10$^{12}$ cm$^{-2}$ and $n_2$=4.0×10$^{12}$ cm$^{-2}$, respectively. Remarkably, the observed $g$-factor enhancement over its bare value predicted by the band theory is as large as $g^*/g_0 = 4.8$ at $n_2$. Although at a still moderate density, such enhancement and its density dependence are more pronounced than what have been reported for other two-dimensional layered materials including $g^*/g_0 < 1.5$ in black phosphorus[37, 38, 39] and $g^*/g_0 \approx 2$ in monolayer WSe$_2$[40]. In conventional 2DEGs, extremely low densities much be approached in order to observe similar (but in fact still less) enhancement, for example, 3 times enhancement of g-factor in GaAs[25] at 0.8×10$^{10}$ cm$^{-2}$, 4.7 times enhancement of $g^*m^*$ in Si inversion layers[24] at 1.0×10$^{11}$ cm$^{-2}$ , and 4.3 times enhancement of $g^*m^*$ in AlAs quantum well[41] at 2.0×10$^{11}$ cm$^{-2}$.

In summary, we observe that the predominant QH states of $\Gamma$ valley holes in $p$-type few-layer WSe$_2$ switch from odd- to even- and then back to odd-integers with reducing the density. By tilting the magnetic field to induce LL crossings, we show that it is the strong Coulomb interaction that enhances $g^*m^*$, a measure of spin susceptibility or $E_z/E_c$, from $5.34 m_0$ to $7.10 m_0$ as the density decreases from 5.7 to 4.0×10$^{12}$ cm$^{-2}$, thereby giving rise to the even-odd alternation. Unprecedentedly, this amounts to a 4.8 times enhancement of the $g$-factor over its bare value predicted by the band theory at the moderate density 4.0×10$^{12}$ cm$^{-2}$. Our findings demonstrate that $p$-type few-layer WSe$_2$ is a superior platform for exploring strongly correlated physics of 2DEGs. For instance, it is promising to observe the Wigner crystallization, long thought to be elusive, at a still moderate density ~$3\times10^{11}$ cm$^{-2}$. This critical density can be further raised if we choose a proper thickness of WSe$_2$, as $m^*$ of $\Gamma$ valley holes ranges from $0.5m_0$ in the bulk to $2.8m_0$ in the monolayer[16, 20] (see Supplementary Information).

**Methods**

WSe$_2$ crystals were grown by chemical vapour transport. Few-layer WSe$_2$ flakes were mechanically exfoliated on SiO$_2$/Si substrates by the scotch tape method. The top h-BN was exfoliated on a PMMA membrane and used for picking up WSe$_2$ flakes. The h-BN/WSe$_2$ structure was then transferred onto a bottom h-BN, previously exfoliated and selected on a 300 nm SiO$_2$/Si substrate. The size of the WSe$_2$ flake was selected such that the flake can be fully encapsulated by the top and bottom h-BN. The encapsulated structures allow us to safely anneal the samples at up to 400 °C in Ar atmosphere to remove small bubbles. To fabricate the metal electrodes, the selective etching process was employed. Standard Hall bar device structure was patterned on a PMMA first, which was also used as the hard mask in the etching process. To expose the contact area of WSe$_2$, reactive ion etching was used for etching the top h-BN. It is possible to control the etching process with 40 sccm O$_2$ as the recipe, because the etching rate of WSe$_2$ is much lower than that of h-BN. Finally, Pd/Au (20/80 nm) was used as the p-type contact via a standard e-beam evaporation.

The transport measurements were conducted using the standard lock-in technique at 1.7 K. In order to study the quantum transport at tilted magnetic fields, we employed a home-built probe with a sample stage that can be *in situ* rotated via a mechanical knob. The tilt angles were internally calibrated by the Hall measurements in order to guarantee the accuracy.

**Acknowledgements**

We acknowledge the financial support from the Research Grants Council of Hong Kong (Projects No. 16302215, No. HKU9/CRF/13G) and the UT-Dallas Research Enhancement Funds. We are grateful for the technical support from the Raith-HKUST Nanotechnology Laboratory for the electron-beam lithography facility at MCPF. F.Z. acknowledges insightful discussions with A. H. MacDonald.

**Author contributions**

N.W., F.Z., and S.X. conceived the projects. S.X. grew the crystal. S.X., L.A. fabricated and measured the devices. S.X., L.A., F.Z. and N.W. analysed the data. F.Z. and N.W. were the principal investigators. S.X., F.Z. and N.W. wrote the manuscript. Other authors provided technical assistances, discussions and comments.

**References**

1. Spivak B, Kravchenko SV, Kivelson SA, Gao XPA. Colloquium. *Rev Mod Phys* 2010, **82**(2): 1743-1766.

2. Wang Z, Shan J, Mak KF. Valley- and spin-polarized Landau levels in monolayer WSe$_2$. *Nat Nano* 2017, **12**(2): 144-149.

3. Ugeda MM, Bradley AJ, Shi S-F, da Jornada FH, Zhang Y, Qiu DY, *et al.* Giant bandgap renormalization and excitonic effects in a monolayer transition metal dichalcogenide semiconductor. *Nat Mater* 2014, **13**(12): 1091-1095.

4. He K, Kumar N, Zhao L, Wang Z, Mak KF, Zhao H, *et al.* Tightly Bound Excitons in Monolayer WSe$_2$. *Phys Rev Lett* 2014, **113**(2): 026803.

5. Zhu B, Chen X, Cui X. Exciton Binding Energy of Monolayer WS$_2$. *Sci Rep* 2015, **5**: 9218.

6. Riley JM, MeevasanaW, BawdenL, AsakawaM, TakayamaT, EknapakulT, *et al.* Negative electronic compressibility and tunable spin splitting in WSe$_2$. *Nat Nano* 2015, **10**(12): 1043-1047.

7. Mak KF, He K, Lee C, Lee GH, Hone J, Heinz TF, *et al.* Tightly bound trions in monolayer MoS$_2$. *Nat Mater* 2013, **12**(3): 207-211.

8. You Y, Zhang X-X, Berkelbach TC, Hybertsen MS, Reichman DR, Heinz TF. Observation of biexcitons in monolayer WSe$_2$. *Nat Phys* 2015, **11**(6): 477-481.

9. Ye Z, Cao T, O'Brien K, Zhu H, Yin X, Wang Y, *et al.* Probing excitonic dark states in single-layer tungsten disulphide. *Nature* 2014, **513**(7517): 214-218.

10. Berkelbach TC, Hybertsen MS, Reichman DR. Theory of neutral and charged excitons in monolayer transition metal dichalcogenides. *Phys Rev B* 2013, **88**(4): 045318.

11. Cudazzo P, Tokatly IV, Rubio A. Dielectric screening in two-dimensional insulators: Implications for excitonic and impurity states in graphane. *Phys Rev B* 2011, **84**(8): 085406.

12. Allain A, Kang J, Banerjee K, Kis A. Electrical contacts to two-dimensional semiconductors. *Nat Mater* 2015, **14**(12): 1195-1205.

13. Cui X, Lee G-H, Kim YD, Arefe G, Huang PY, Lee C-H, *et al.* Multi-terminal transport measurements of MoS$_2$ using a van der Waals heterostructure device platform. *Nat Nano* 2015, **10**(6): 534-540.

14. Xu S, Wu Z, Lu H, Han Y, Long G, Chen X, *et al.* Universal low-temperature Ohmic contacts for quantum transport in transition metal dichalcogenides. *2D Mater* 2016, **3**(2): 021007.

15. Wu Z, Xu S, Lu H, Khamoshi A, Liu G-B, Han T, *et al.* Even–odd layer-dependent magnetotransport of high-mobility Q-valley electrons in transition metal disulfides. *Nat Commun* 2016, **7**: 12955.

16. Xu S, Shen J, Long G, Wu Z, Bao Z-q, Liu C-C, *et al.* Odd-Integer Quantum Hall States and Giant Spin Susceptibility in *p*-Type Few-Layer WSe$_2$. *Phys Rev Lett* 2017, **118**(6): 067702.

17. Fallahazad B, Movva HCP, Kim K, Larentis S, Taniguchi T, Watanabe K, *et al.* Shubnikov-de Haas Oscillations of High-Mobility Holes in Monolayer and Bilayer WSe$_2$: Landau Level Degeneracy, Effective Mass, and Negative Compressibility. *Phys Rev Lett* 2016, **116**(8): 086601.

18. Movva HCP, Fallahazad B, Kim K, Larentis S, Taniguchi T, Watanabe K, *et al.* Density-Dependent Quantum Hall States and Zeeman Splitting in Monolayer and Bilayer $WSe_2$. *Phys Rev Lett* 2017, **118**(24): 247701.

19. Tanatar B, Ceperley DM. Ground state of the two-dimensional electron gas. *Phys Rev B* 1989, **39**(8): 5005-5016.

20. Wickramaratne D, Zahid F, Lake RK. Electronic and thermoelectric properties of few-layer transition metal dichalcogenides. *J Chem Phys* 2014, **140**(12): 124710.

21. Geick R, Perry CH, Rupprecht G. Normal Modes in Hexagonal Boron Nitride. *Phys Rev* 1966, **146**(2): 543-547.

22. Wang Z, Mak KF, Shan J. Strongly interaction-enhanced valley magnetic response in monolayer $WSe_2$. *arXiv preprint* 2017: arXiv:1705.01078.

23. Okamoto T, Hosoya K, Kawaji S, Yagi A. Spin Degree of Freedom in a Two-Dimensional Electron Liquid. *Phys Rev Lett* 1999, **82**(19): 3875-3878.

24. Pudalov VM, Gershenson ME, Kojima H, Butch N, Dizhur EM, Brunthaler G, *et al.* Low-Density Spin Susceptibility and Effective Mass of Mobile Electrons in Si Inversion Layers. *Phys Rev Lett* 2002, **88**(19): 196404.

25. Tutuc E, Melinte S, Shayegan M. Spin Polarization and *g* Factor of a Dilute GaAs Two-Dimensional Electron System. *Phys Rev Lett* 2002, **88**(3): 036805.

26. Zhu J, Stormer HL, Pfeiffer LN, Baldwin KW, West KW. Spin Susceptibility of an Ultra-Low-Density Two-Dimensional Electron System. *Phys Rev Lett* 2003, **90**(5): 056805.

27. Shkolnikov YP, Vakili K, De Poortere EP, Shayegan M. Dependence of Spin Susceptibility of a Two-Dimensional Electron System on the Valley Degree of Freedom. *Phys Rev Lett* 2004, **92**(24): 246804.

28. Padmanabhan M, Gokmen T, Bishop NC, Shayegan M. Effective Mass Suppression in Dilute, Spin-Polarized Two-Dimensional Electron Systems. *Phys Rev Lett* 2008, **101**(2): 026402.

29. Wang L, Meric I, Huang PY, Gao Q, Gao Y, Tran H, *et al.* One-Dimensional Electrical Contact to a Two-Dimensional Material. *Science* 2013, **342**(6158): 614.

30. Chen X, Wu Y, Wu Z, Han Y, Xu S, Wang L, *et al.* High-quality sandwiched black phosphorus heterostructure and its quantum oscillations. *Nat Commun* 2015, **6:** 7315.

31. Li X, Zhang F, Niu Q. Unconventional Quantum Hall Effect and Tunable Spin Hall Effect in Dirac Materials: Application to an Isolated $MoS_2$ Trilayer. *Phys Rev Lett* 2013, **110**(6): 066803.

32. Nicholas RJ, Haug RJ, Klitzing Kv, Weimann G. Exchange enhancement of the spin splitting in a $GaAs-Ga_xAl_{1-x}As$ heterojunction. *Phys Rev B* 1988, **37**(3): 1294-1302.

33. Fang FF, Stiles PJ. Effects of a Tilted Magnetic Field on a Two-Dimensional Electron Gas. *Phys Rev* 1968, **174**(3): 823-828.

34. Kurganova EV, van Elferen HJ, McCollam A, Ponomarenko LA, Novoselov KS, Veligura A, *et al.* Spin splitting in graphene studied by means of tilted magnetic-field experiments. *Phys Rev B* 2011, **84**(12)**:** 121407.

35. Cao Y, Mishchenko A, Yu GL, Khestanova E, Rooney AP, Prestat E, *et al.* Quality Heterostructures from Two-Dimensional Crystals Unstable in Air by Their Assembly in Inert Atmosphere. *Nano Lett* 2015, **15**(8)**:** 4914-4921.

36. Roth LM, Lax B, Zwerdling S. Theory of Optical Magneto-Absorption Effects in Semiconductors. *Phys Rev* 1959, **114**(1)**:** 90-104.

37. Li L, Yang F, Ye GJ, Zhang Z, Zhu Z, Lou W, *et al.* Quantum Hall effect in black phosphorus two-dimensional electron system. *Nat Nano* 2016, **11**(7)**:** 593-597.

38. Long G, Maryenko D, Shen J, Xu S, Hou J, Wu Z, *et al.* Achieving Ultrahigh Carrier Mobility in Two-Dimensional Hole Gas of Black Phosphorus. *Nano Lett* 2016, **16**(12)**:** 7768-7773.

39. Tran S, Yang J, Gillgren N, Espiritu T, Shi Y, Watanabe K, *et al.* Surface transport and quantum Hall effect in ambipolar black phosphorus double quantum wells. *Sci Adv* 2017, **3**(6)**:** e1603179.

40. Martin V. Gustafsson MY, Carlos Forsythe, Daniel Rhodes, Kenji Watanabe, Takashi Taniguchi, James Hone, Xiaoyang Zhu, Cory R. Dean. Ambipolar Landau levels and strong exchange-enhanced Zeeman energy in monolayer WSe2. *arXiv preprint* 2017**:** arXiv:1707.08083.

41. Gokmen T, Padmanabhan M, Tutuc E, Shayegan M, De Palo S, Moroni S, *et al.* Spin susceptibility of interacting two-dimensional electrons with anisotropic effective mass. *Phys Rev B* 2007, **76**(23)**:** 233301.

**Figures:**

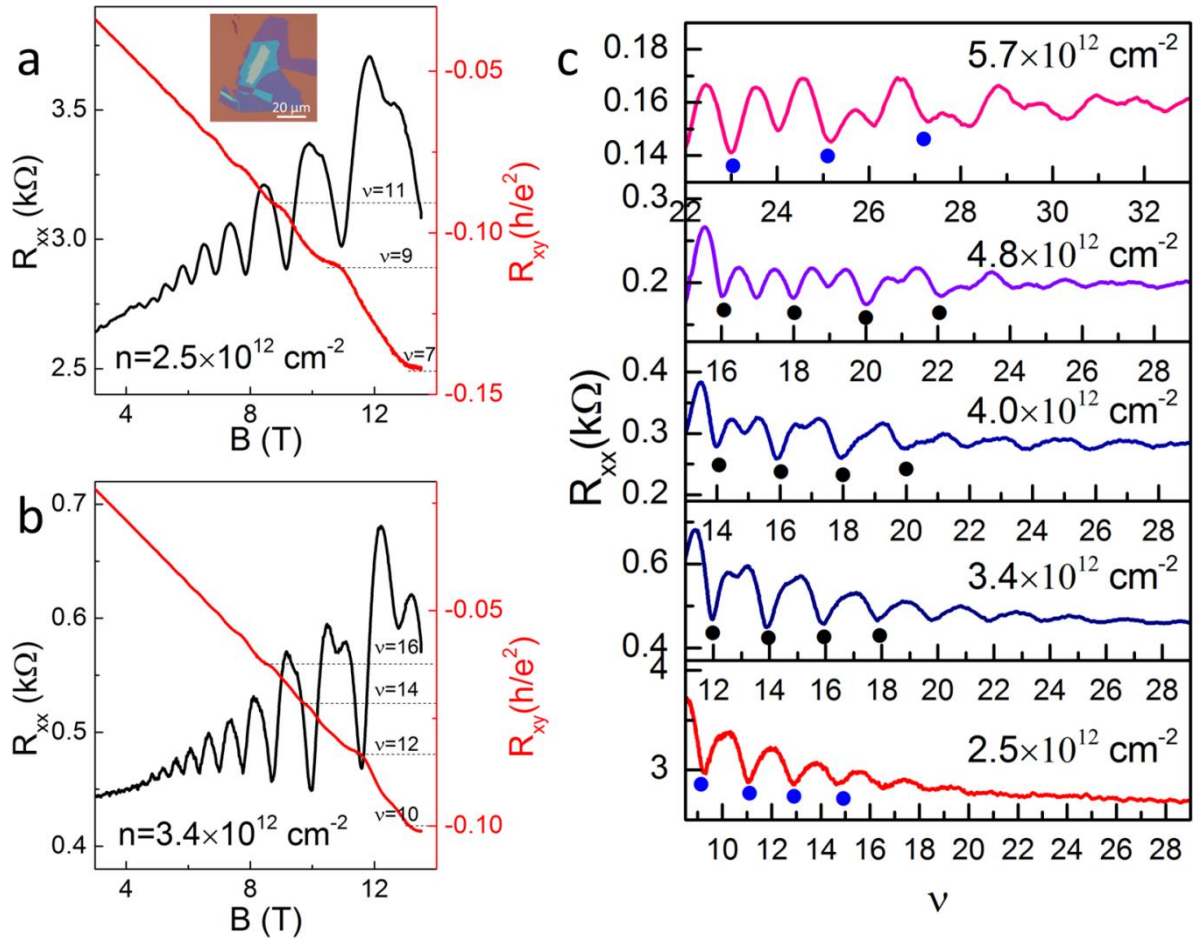

**Figure 1: Quantum oscillations at various different carrier densities.** $R_{xx}$ and $R_{xy}$ plotted as functions of $B$ at (**a**) $n = 2.5 \times 10^{12}$ cm$^{-2}$ and (**b**) $n = 3.4 \times 10^{12}$ cm$^{-2}$. QH plateaus in $R_{xy}$ are marked by dashed lines. The inset in (**a**) is an optical image of the h-BN encapsulated WSe$_2$ structure. (**c**) $R_{xx}$ plotted as functions of $\nu$ at various different $n$. The FFs are calculated by $\nu = nh/eB$. The main minima are marked by solid dots. The predominant QH states at odd- and even-integer FFs are distinguished by blue and black dots, respectively. All data were recorded at T=1.7 K.

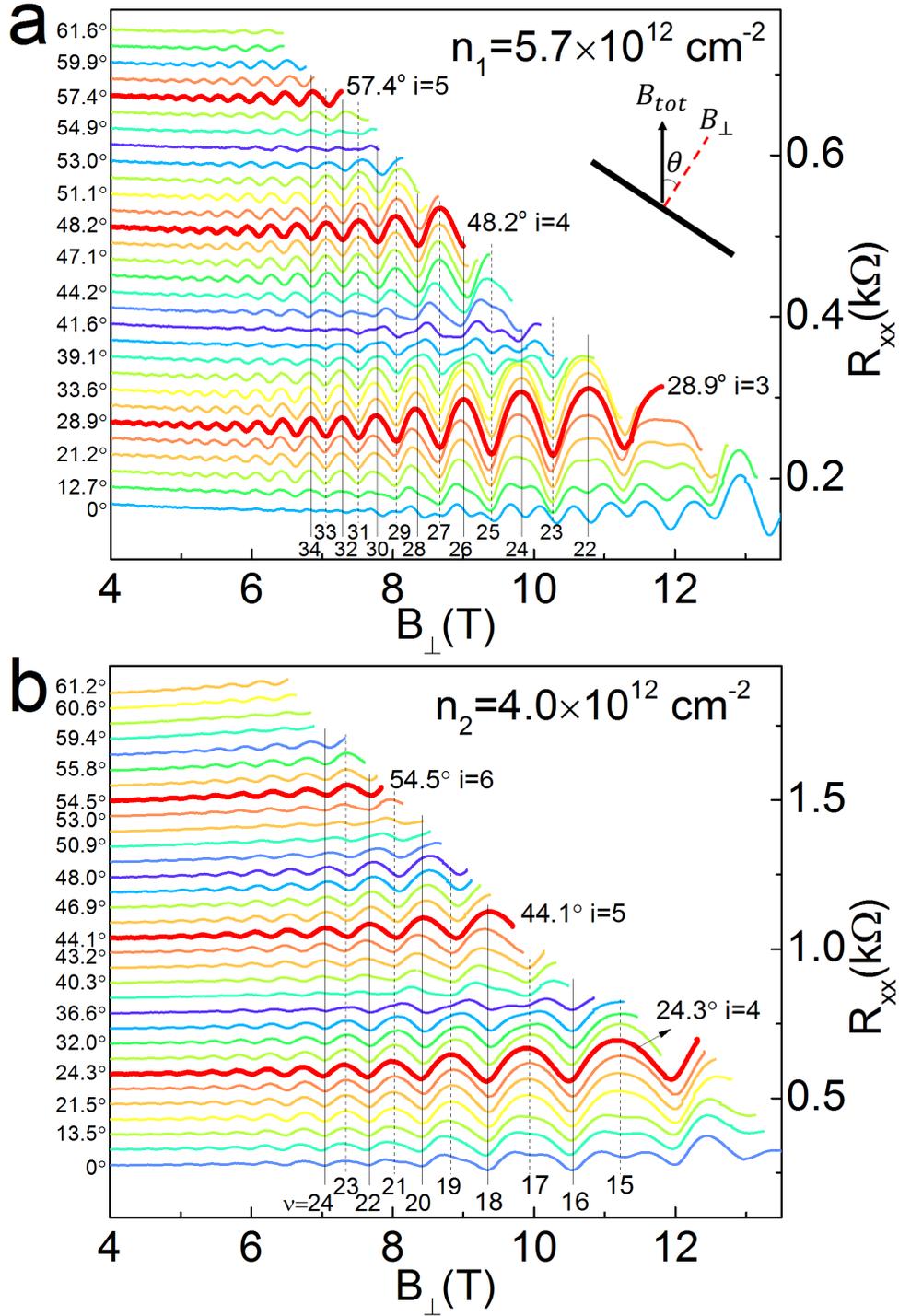

**Figure 2: Density-dependent Landau level crossings at tilted magnetic fields.** $R_{xx}$ plotted as functions of $B_\perp$ at various different tilt angles for two representative densities: (a) $n_1 = 5.7 \times 10^{12}$ cm$^{-2}$, (b) $n_2 = 4.0 \times 10^{12}$ cm$^{-2}$. The $R_{xx}$ curves at the coincident angles are labelled and marked in red. The data at different tilt angles are vertically offset for clarity. The tilt angle is defined in the inset of (a).

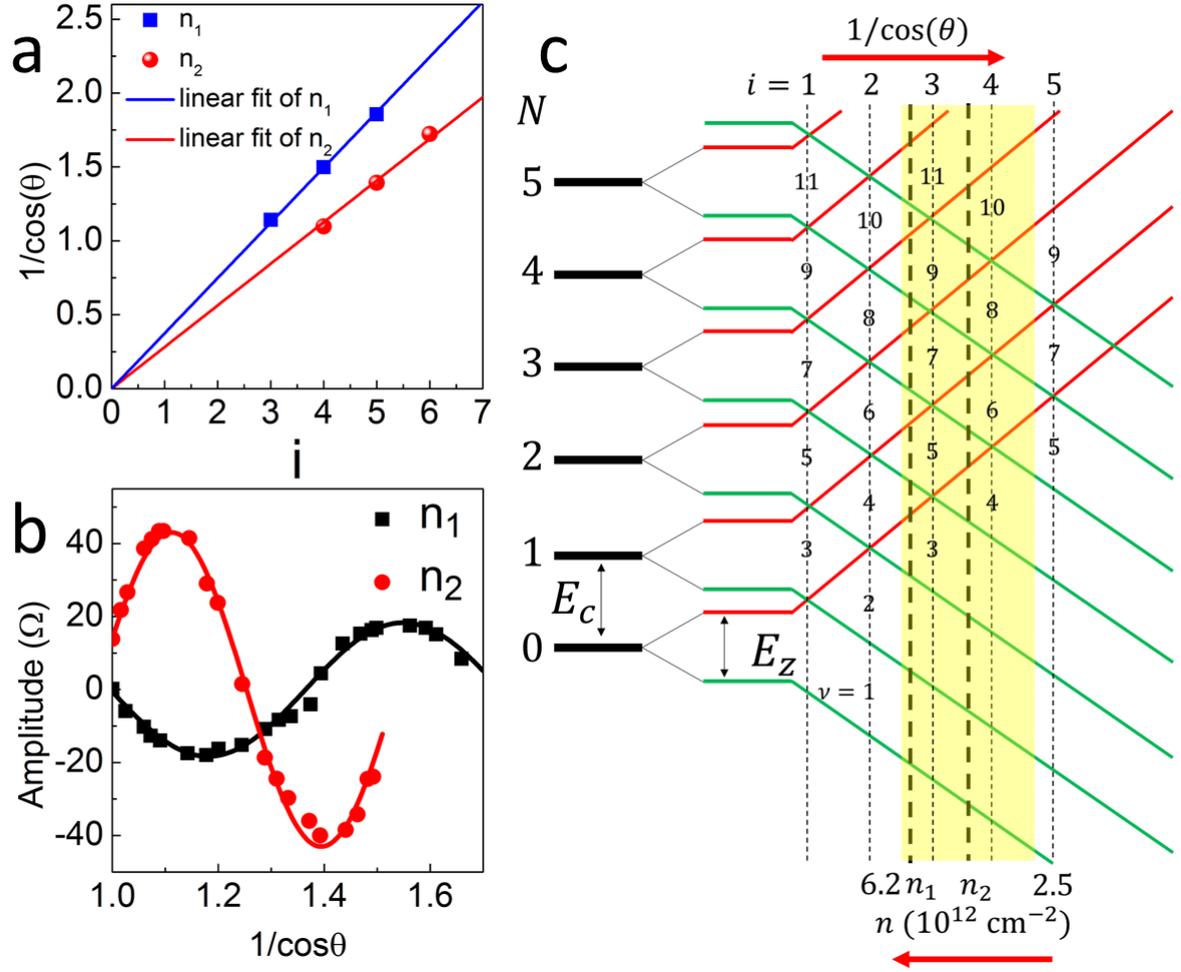

**Figure 3: Determination of the density-dependent spin susceptibility**. (**a**) The values of $1/cos\theta$ at coincident angles plotted as functions of the coincidence index $i$ for $n_1$ and $n_2$. The slopes from the linear fitting yield the values of $g^*m^*$. (**b**) SdH amplitudes plotted as functions of $1/cos\theta$ for $n_1$ (recorded at $\nu = 29$) and $n_2$ (recorded at $\nu = 19$). The data fitting to cosine functions yield the values of $g^*m^*$. (**c**) A schematic diagram depicts the LL evolution with the tilt angle $\theta$ or the carrier density $n$. Under a fixed $B_\perp$, $E_c$ does not change while $E_z$ scales linearly with $B_{\text{tot}}$; $E_z/E_c$ can be enhanced by increasing $\theta$ or by decreasing $n$. The yellow strip marks the density region probed in our experiment, with the bold dashed lines denoting the locations of $\theta = 0°$ for $n_1$ and $n_2$.

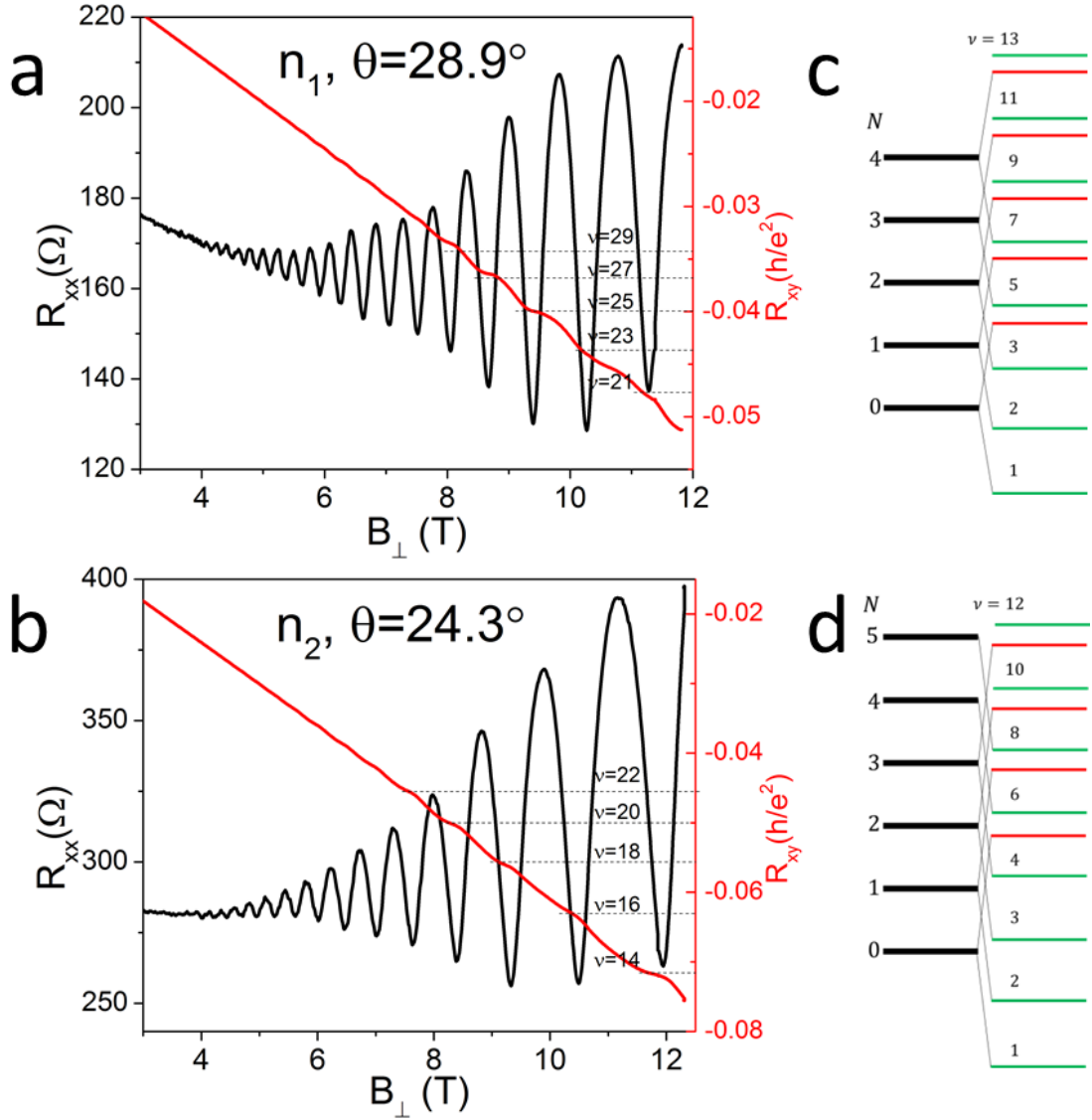

**Figure 4: Quantum Hall states at the first coincident angles.** $R_{xx}$ and $R_{xy}$ plotted as functions of $B_\perp$ at the first coincident angles for (**a**) $n_1 = 5.7 \times 10^{12}$ cm$^{-2}$ ($\theta = 28.9°$) and (**b**) $n_2 = 4.0 \times 10^{12}$ cm$^{-2}$ ($\theta = 24.3°$). The QH plateaus in $R_{xy}$ show odd- and even-integer sequences in (**a**) and (**b**), respectively. (**c**) and (**d**) are schematics of LLs at zero tilt angles for $n_1$ and $n_2$, respectively. The green and red lines distinguish the spin-down and spin-up states.